 \documentclass[final,5p,times,twocolumn]{elsarticle}

\usepackage{amssymb}
\usepackage{lipsum}
\usepackage{graphicx}
\usepackage{subcaption}

\journal{Nuclear Physics A}

\begin{document}

\begin{frontmatter}



\title{Machine Learning for the Cluster Reconstruction in the CALIFA Calorimeter at R3B}

\author[1]{T. Jenegger}
\author[1]{N. Hartman}
\author[1]{R. Gernh\"auser}
\author[1]{L. Fabbietti}
\author[1]{L. Heinrich}
\address[1]{{TUM School of Natural Sciences, Technical University of Munich},
            {Germany}}

\begin{abstract}

The R3B experiment at FAIR studies nuclear reactions using high-energy radioactive beams. One key detector in R3B is the CALIFA calorimeter consisting of 2544 CsI(Tl) scintillator crystals designed to detect light charged particles and  gamma rays with an energy resolution in the per cent range after Doppler correction. Precise cluster reconstruction from sparse hit patterns is a crucial requirement. Standard algorithms typically use fixed cluster sizes or geometric thresholds. To enhance performance, advanced machine learning techniques such as agglomerative clustering were implemented to use the full multi-dimensional parameter space including geometry, energy and time of individual interactions. An Edge Detection Neural Network exhibited significant differences. This study, based on Geant4 simulations, demonstrates improvements in cluster reconstruction efficiency of more than 30\%, showcasing the potential of machine learning in nuclear physics experiments.
\end{abstract}




\end{frontmatter}




\section{Introduction}
\label{sec:intro}
With the advancements in facilities dedicated to the production of radioactive beams at relativistic energies, such as the Facility for Antiproton and Ion Research (FAIR) at GSI, significant progress is expected for our understanding of exotic nuclei far from stability \cite{kalantar2024experiments}. FAIR will provide high-intensity relativistic radioactive beams of rare isotopes with energies in the range of 1 A GeV, enabling investigations with full kinematic reconstruction \cite{leifels2025status}.
A key experimental setup designed for this purpose is the \textbf{R}eactions with \textbf{R}elativistic \textbf{R}adioactive \textbf{B}eams (R3B) setup, providing access to high-resolution spectroscopic data. This setup serves as a unique tool for unveiling the structure of nuclei and their reaction dynamics with unprecedented precision.\newline
At the core of the R3B Setup is the CALIFA calorimeter (Calorimeter for the In-Flight Detection of Gamma Rays and Light Charged Particles), a highly segmented detection system composed of 2544 CsI(Tl) scintillator crystals that hermetically enclose the target area in the polar angular range of $7^\circ < \theta < 140^\circ$. This design enables the simultaneous measurement of gamma rays down to $E_{\gamma} \approx$ 100 keV and light charged particles, such as protons and deuterons, up to several hundred A MeV \cite{cortina2014califa}. To ensure optimal performance, extensive research has been conducted to refine the geometric design, minimize scattering and energy loss due to the mechanical structure \cite{alvarez2014performance}, and develop a dead-time-free data acquisition system capable of handling high-rate experiments \cite{ledigital}. Furthermore, a seamless integration within the R3BRoot framework \cite{bertini2011r3broot} has been achieved, enabling offline data analysis from the raw to the calibrated data level and ultimately to the cluster level, where individual hits are recombined for the final energy reconstruction.\newline
This study presents the results of a hierarchical machine learning model to enhance the energy reconstruction of gamma rays in CALIFA. Using simulated Geant4 data, the performance of the geometrical R3B clustering algorithm is compared to an agglomerative clustering model \cite{Nielsen2016} and a multi-layer perceptron architecture~\cite{popescu2009multilayer}, demonstrating the potential of machine learning techniques in improving reconstruction efficiency and accuracy.\newline

\section{Methodology}
\label{sec:metho}
\subsection{Challenges in Relativistic Gamma Spectroscopy}\label{s_sec:gamma_spec}
\begin{figure}[!htb]
	\centering 
	\includegraphics[width=0.49\textwidth]{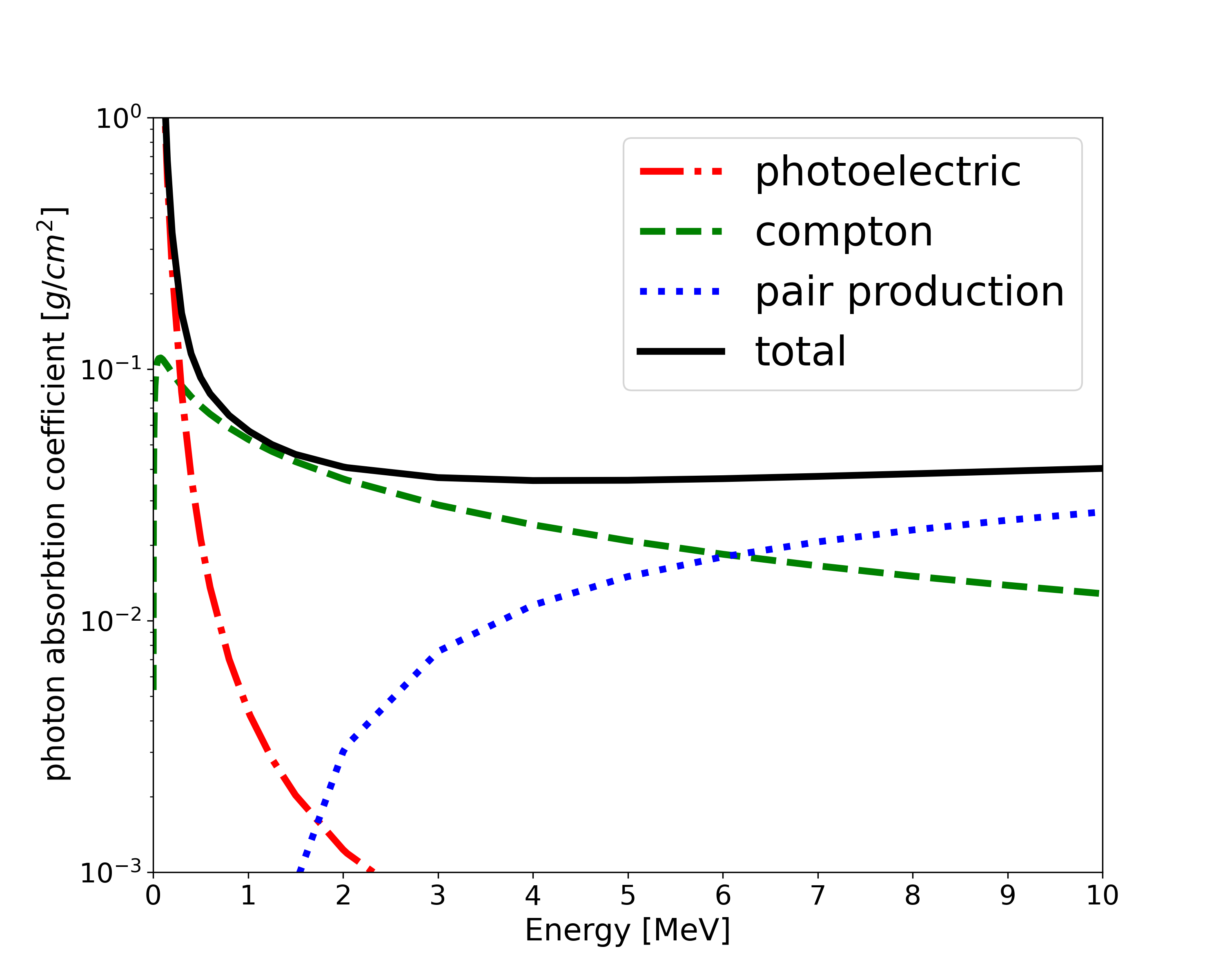}	
	\caption{Photon absorption coefficients in CsI in the range from $100$ keV to $10$ MeV with data from XCOM database \cite{seltzer2010xcom}.} 
	\label{fig:csi}%
\end{figure}

While the detection of light charged particles such as protons typically yields well-localized energy deposits in segmented detector arrays, the detection of gamma rays which emerge from the reaction vertex presents significant challenges. These primarily arise from the inherently sparse and spatially distributed energy deposits resulting from the interaction mechanisms of photons with the scintillator material (see Fig. \ref{fig:csi}) \cite{kolanoski2016teilchendetektoren}.\newline
At photon energies below approximately $300$ keV, the photoelectric effect dominates the interaction cross-section in the CALIFA detector material (CsI(Tl)). As the photon energy increases, Compton scattering becomes the predominant process. For photon energies exceeding the pair production threshold ($E_{\gamma} > 2m_{e}c^2 \approx 1.022 MeV$), electron-positron pair creation becomes possible and is the dominant interaction mechanism above $E_{\gamma} \approx 6$ MeV.\newline
Compton scattering broadens the clustering by the deflection of the incident gamma ray. According to the Klein-Nishina formula, the scattering is predominantly forward-focused for moderate to high photon energies \cite{klein1929streuung}, leading to additional clusters in neighboring crystals.\newline
At high photon energies, the dominant interaction mechanism in the detector material is pair production (see Fig.~\ref{fig:csi}), in which the incident photon converts into an electron-positron pair during the initial interaction. The subsequent annihilation of the positron results in the emission of two additional gamma photons, each with an energy of $511$ keV. These secondary photons often escape the initial interaction site, leading to a significant fraction of the incident photon’s energy being deposited in multiple detector elements.\newline
For gamma rays emitted by nuclei at rest, this behavior gives rise to well-defined single- and double-escape peaks in the recorded energy spectra -- corresponding to the escape of one or both $511\,\mathrm{keV}$ photons, respectively -- if these photons exit the cluster volume without interaction.\newline
In experiments involving relativistic ion beams, such as those exploited at R3B, Doppler broadening significantly affects the observed spectral features, including the single- and double-escape peaks. Moreover, for primary gamma rays with energies well above the pair production threshold ($E_{\gamma} > 2m_{e}c^2$), both the electron and the positron produced in the initial interaction are subject to substantial energy loss via Bremsstrahlung. These effects contribute to a complex and highly non-trivial interaction pattern of gamma rays within the segmented detector system.\newline

\subsection{Data Structure and geometrical R3B Clustering Algorithm}\label{s_sec:r3b_clustering}
The fundamental data entity in the analysis is a hit, defined as a discrete signal recorded by an individual detector segment at a specific time. To suppress contributions from low-energy background, only signals exceeding a predefined energy threshold are registered. In the present analysis, this threshold was set to 100 keV.\newline
In the standard data acquisition (DAQ) configuration, all CALIFA detector hits occurring within a $\pm 4\,\mu\mathrm{s}$ time window are grouped into a single event. Each individual hit \( i \) in one of the detector crystals is represented by a data structure containing the calibrated energy deposit \( E_i \), the polar angle \( \theta_i \), the azimuthal angle \( \phi_i \), and a time stamp \( t_i \), which is synchronized using the White Rabbit Precision Time Protocol~\cite{lipinski2011white}.\newline
In the geometrical R3B clustering approach, the time information $t_i$ is not utilized during the spatial reconstruction of clusters.\newline
The initial stage of the clustering algorithm begins by sorting all hits in descending order of energy. A user-defined geometric condition, typically a cone emerging from the central target point with an aperture of $0.25\,\mathrm{rad}$, is applied. This value has been found to provide an optimal compromise between compact high-energy clusters from light charged particles and more diffuse gamma-ray showers.\newline
The hit with the highest energy defines the seed or center of the first cluster. The algorithm then iterates through the remaining hits and includes each hit to the current cluster if it sits within the specified cone relative to the seed direction. Once the list is fully processed for the current cluster, the next highest-energy unassigned hit becomes the seed of a new cluster. This procedure repeats until no unassigned hits remain.\newline
\subsection{Simulation Setup}\label{s_sec:data_sim}
Simulated datasets are used to evaluate and compare the performance of the clustering algorithms presented in this work.
A geometrical model of the detector, closely matching the experimental setup, was implemented within the R3BROOT framework. The simulation employs a GEANT4-based Monte Carlo~\cite{agostinelli2003geant4} back-end, which accounts for all relevant secondary interaction processes. This approach enables realistic modeling of energy deposition and provides access to ground-truth labels for each individual interaction.\newline

The CALIFA detector geometry used in the simulation corresponds to the configuration implemented in early 2024. At that time, the iPhos region (polar angles $19^\circ$ -- $43^\circ$) was fully instrumented, while only the forward half of the Barrel region ($43^\circ$ -- $87^\circ$) was active. The forward-most CEPA region ($7^\circ$ -- $19^\circ$) was not yet equipped.\newline
Gamma-ray energies were sampled from a uniform distribution between $0.3\,\mathrm{MeV}$ and $10\,\mathrm{MeV}$. The interaction of the primary gamma rays with the CsI(Tl) scintillation material was modeled using Geant4.\newline
To emulate realistic event topologies of signal and background, three gamma rays were generated per event, resulting in multiple detector hits. Timing information was coarsely simulated by assigning to each primary gamma a random emission time within the $\pm 4\,\mu\mathrm{s}$ event window. The corresponding hit times were then Gaussian-smeared with a standard deviation of $200\,\mathrm{ns}$ to reflect the timing spread of the electronic signal of slow CsI(Tl) scintillator crystals.\newline
Event selection is limited to cases in which all three gamma rays are---at least partially---detected within the geometrical acceptance of the CALIFA detector, which only partially encloses the target region. For gamma rays that deposit only a fraction of their energy in the detector volume, the corresponding true energy is adjusted to reflect only the energy actually deposited in CALIFA.\newline
The resulting dataset was split into training and test subsets, comprising 13{,}000 and 7{,}000 events, respectively.

\subsection{Performance Metrics}\label{s_sec:metrics}
To quantitatively assess the performance of the clustering algorithms presented in this work, a set of four custom metrics was defined. Three of these are event-based, while an optional fourth metric evaluates clustering quality on a per-cluster basis:
\begin{itemize}
    \item \textbf{True Positive (TP)}: All hits in an event are correctly assigned to their respective clusters.
    \item \textbf{False Positive (FP)}: At least one hit in an event is incorrectly merged into a cluster it does not belong to.
    \item \textbf{False Negative (FN)}: At least one hit is not merged into its true cluster and instead forms a spurious cluster.
    \item \textbf{False Mixed (FM)}: An event is classified as false mixed if it contains both FP and FN characteristics -- i.e., at least one hit is incorrectly merged, and at least one true cluster is partially reconstructed.
\end{itemize}
In addition, a cluster-based metric is defined:
\begin{itemize}
    \item \textbf{Well Reconstructed (WR)}: The ratio of correctly reconstructed clusters to the total number of true clusters in the dataset.
\end{itemize}
These metrics allow a comprehensive evaluation of clustering accuracy, robustness, and failure modes.\newline
Special attention must be given to the false negative rate, which is closely associated with the complex interaction pattern in the segmented detector. These processes produce widely spread hits that cannot be merged using the geometrical R3B clustering method, thereby motivating the development of a multi-layer perceptron architecture to improve clustering performance at the boundaries (see Subsection \ref{s_sec:edge}).\newline

\subsection{Agglomerative Clustering}\label{s_sec:agglo}
To incorporate temporal information into the clustering process---unlike the geometrical R3B algorithm, which omits it---a generic, well-established method was adopted: agglomerative clustering \cite{Nielsen2016} as implemented in the \texttt{SciPy} library \cite{virtanen2020scipy}. This unsupervised learning algorithm enables flat clustering based on hierarchical linkage with a user-defined threshold.\newline
Each hit was mapped into spherical coordinates \((\theta, \phi, r)\), where the radial component \(r\) encodes time information. To ensure non-negative radii, the acquisition time window of \(\pm 4\,\mu\mathrm{s}\) was shifted by \(+4.5\,\mu\mathrm{s}\). The Ward linkage criterion \cite{nielsen2016hierarchical}, which minimizes intra-cluster variance, was employed as the distance metric.\newline
The threshold parameter was optimized to yield the best performance according to the custom-defined \textit{true positive} (TP) and \textit{well reconstructed} (WR) metrics.\newline
As shown in Table~\ref{tab:results}, the agglomerative clustering algorithm demonstrates improved performance both on an event level (true positive rate) and on a cluster level (correctly reconstructed clusters) compared to the geometrical R3B clustering. However, this improvement is accompanied by an increased false negative rate, indicating that the algorithm tends to under-merge hits near the edges of clusters. This limitation motivated the development and application of an edge detection neural network, which is introduced in the following subsection.

\subsection{Edge Detection Neural Network}\label{s_sec:edge}
To enhance the clustering performance, particularly at the boundaries of hit distributions, a multi-layer perceptron architecture was developed using the Pytorch library \cite{imambi2021pytorch} to perform pairwise classification of detector hits. This model is applied either to individual raw hits or to hits pre-clustered via agglomerative clustering, on an event-by-event basis.

The model takes 12 input features for each hit pair $(i, j)$: absolute values of energy $(E_i, E_j)$, polar angle $(\theta_i, \theta_j)$, azimuthal angle $(\phi_i, \phi_j)$, and time $(t_i, t_j)$. Additionally, four differential features are computed: $\Delta E = |E_i - E_j|$, $\Delta \theta = |\theta_i - \theta_j|$, $\Delta \phi = |\phi_i - \phi_j|$, and $\Delta t = |t_i - t_j|$. These differential inputs are helpful for training stability and convergence with our limited model sizes tested. In particular, $\Delta \phi$ resolves the discontinuities caused by the periodicity of the azimuthal angle (e.g., distinguishing between $\phi = 355^\circ$ and $\phi = 5^\circ$), which would otherwise introduce large erroneous differences in angular comparisons.

Of the 12 features, only the hit time is normalized to the $[0, 1]$ interval; all other values are used in their native physical units.
The neural network architecture takes the 12-dimensional input vector and passes it through a fully connected feed-forward network with one hidden layer of $10^3$ nodes, followed by a rectifier linear unit activation function (ReLU)~\cite{agarap2018deep}. Two additional hidden layers, each with $10^2$ nodes, are applied sequentially. The output layer consists of a single node with a sigmoid activation, yielding a score in the interval $[0, 1]$, where values close to 1 indicate that the hits (or clusters) are likely to originate from the same event cluster.\newline

Training is performed using the binary cross-entropy loss function \cite{mannor2005cross,de2005tutorial} and stochastic gradient descent (SGD) \cite{newton2018recent} with a fixed learning rate of $5 \times 10^{-3}$. Given the moderate size of the training dataset, full-batch training is employed without mini-batching. The model is trained for $8 \times 10^4$ epochs. After training, a threshold is applied to the prediction scores to classify hit pairs. This threshold is tuned to optimize the performance across all defined metrics, as described in Subsection~\ref{s_sec:metrics}. Final clusters are then formed by grouping all connected hit pairs based on the predicted associations.

\begin{figure}[!htb]
        \centering
        \includegraphics[width=0.45\textwidth]{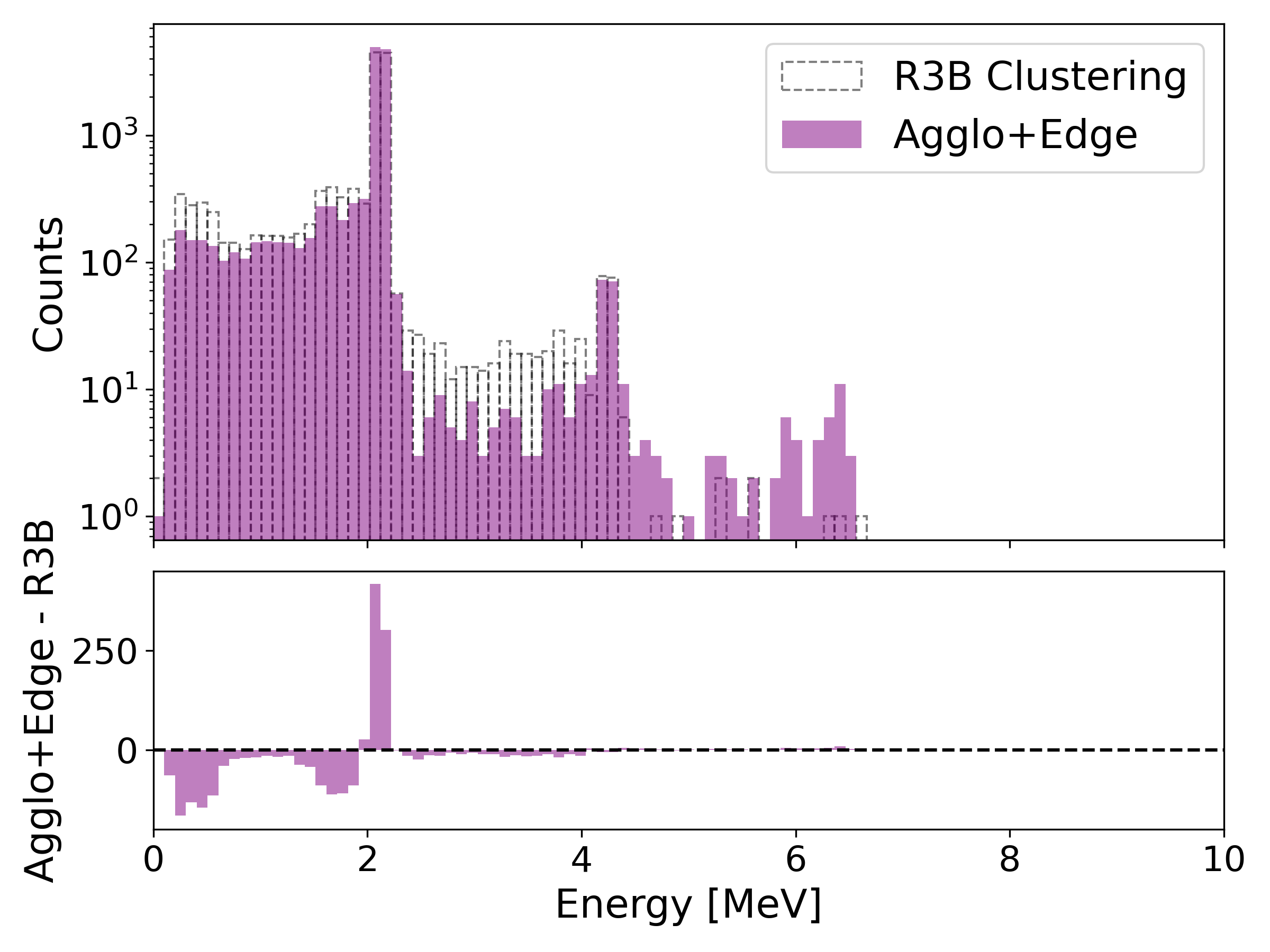}
        \caption{Reconstructed gamma energy spectrum from simulated events, each consisting of three 2.1~MeV gamma photons emitted from the target point. The upper panel shows the comparison between the geometrical R3B clustering and the Agglo+Edge method. The lower panel displays the bin-by-bin count difference between the two approaches. The Agglo+Edge model demonstrates a significant improvement by successfully reattaching escaped hits, notably in cases where sparse energy deposits around 1.6~MeV and 0.5~MeV result from pair production and subsequent annihilation processes of the original gamma photons. This clean-up step leads to a marked reduction in false negatives compared to the geometrical R3B clustering.}
        \label{fig:2_1_mev_r3b_aggloNN}%
\end{figure}

The edge detection NN was implemented and tested in three configurations:

\begin{itemize}
    \item \textbf{Plain Edge NN:} The model is applied directly to individual hits without any pre-clustering. All clustering is performed based solely on the NN predictions.
    \item \textbf{R3B + Edge NN:} The data are first clustered using the geometrical R3B clustering algorithm as an initial clean-up step. For each resulting cluster, an energy-weighted center of mass is calculated, replacing individual hits. The  NN is then trained exclusively on false negative cases, i.e., events where reconstructed clusters exhibit detached hits. In application, the geometrical R3B clustering is first applied to the test data, followed by the NN to refine cluster boundaries and reduce the false negative rate as clean-up step.
    \item \textbf{Agglo + Edge NN:} This strategy mirrors the R3B+Edge approach, with the key difference that time information is incorporated. As in the R3B+Edge model, the NN is trained on false negative cases to perform a final clean-up step after pre-clustering the hits using the agglomerative clustering algorithm described in the previous subsection. The significant reduction of the false negative rate achieved by the clean-up step in the Agglo+Edge implementation is demonstrated in Figure~\ref{fig:2_1_mev_r3b_aggloNN}, which compares the reconstructed energy spectra from simulations of mono-energetic 2.1~MeV gamma events using the geometrical R3B clustering and the Agglo+Edge method.
\end{itemize}
\section{Discussion}\label{sec:results}
\begin{figure}[!htb]
        \centering
        \includegraphics[width=0.45\textwidth]{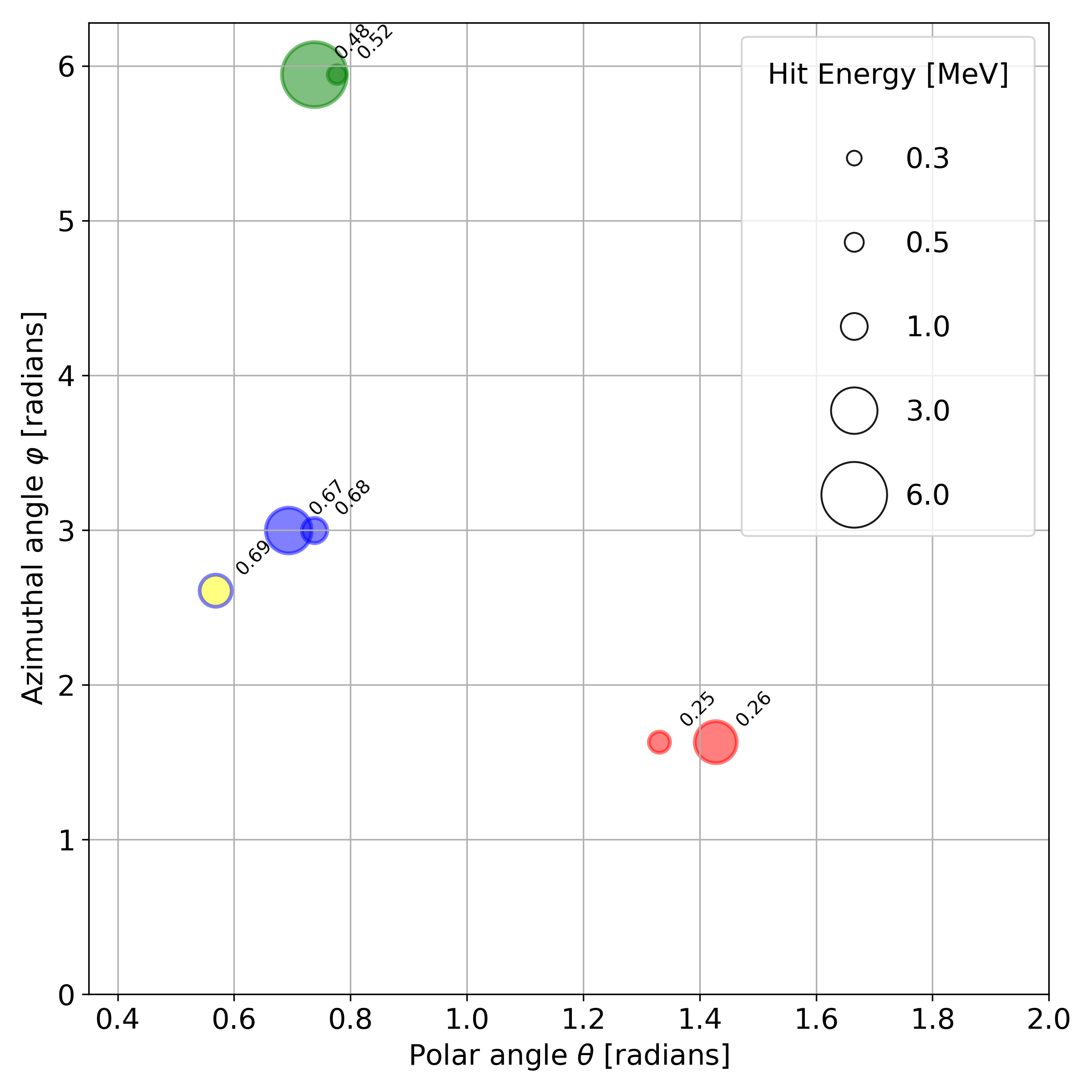}
	    \caption{Example of a simulated event involving three primary gamma photons, illustrating the performance difference between the Agglo+Edge clustering method and the geometrical R3B clustering approach. Each marker represents a detected hit, plotted as a function of the polar angle \(\theta\) and the azimuthal angle \(\varphi\). The edge color of each circle indicates the true cluster assignment (ground truth), while the fill color denotes the cluster assignment according to the geometrical R3B clustering. The size of each circle reflects the energy deposited in the detector segment. Numbers adjacent to the hits represent the normalized hit times. In this event, the geometrical R3B clustering incorrectly assigns the hit at \((\theta \approx 0.6\,\mathrm{rad},\, \varphi \approx 2.7\,\mathrm{rad})\), with a normalized time of 0.69 (blue edge, yellow fill), to a separate cluster, resulting in a \textit{False Negative} (FN). In contrast, the Agglo+Edge method correctly assigns all hits to their respective clusters.}
        \label{fig:comparison_r3b_agglo_edge}%
\end{figure}

\begin{table}[h!]
\begin{center}
\resizebox{0.5\textwidth}{!}{%
\begin{tabular}{||c| c| c |c | c|| c|}
 \hline
 Clustering Model & TP($\uparrow$) & FP($\downarrow$) & FN($\downarrow$) & FM($\downarrow$) & WR($\uparrow$) \\ [0.5ex] 
 \hline\hline
 Geometrical R3B Clustering & 60.6 & 5.3 & 25.2 & 8.9 & 80.4 \\ 
 \hline
 Agglomerative Clustering & 62.8 & \textbf{3.3} & 32.0 & 1.9 & 84.1 \\ 
 \hline
 Edge Clustering (no time) & 63.4$\pm$0.3 & 7.2$\pm$0.3 & 24.8$\pm$0.7 & 4.6$\pm$0.1 & 82.4$\pm$0.1 \\ 
 \hline
 Edge Clustering (with time) & 74.7$\pm$0.5 & 3.4$\pm$0.6 & 20.5$\pm$1.3 & 1.4$\pm$0.1 & 89.2$\pm$0.1 \\ 
 \hline
 R3B + Edge (no time) & 67.4$\pm$0.3 & 8.5$\pm$0.3 & 16.0$\pm$0.4 & 8.0$\pm$0.3 & 82.2$\pm$0.1 \\ 
 \hline
 Agglo + Edge (with time) & \textbf{81.3$\pm$0.3} & 5.1$\pm$0.0 & \textbf{12.2$\pm$0.3} & \textbf{1.5$\pm$0.1} & \textbf{91.0$\pm$0.1} \\
 \hline
\end{tabular}
}
\end{center}
\caption{Summary of performance metrics as defined in Subsection~\ref{s_sec:metrics}, evaluated for the different clustering algorithms. The models \textit{Geometrical R3B Clustering}, \textit{Edge Clustering (no time)}, and \textit{R3B + Edge (no time)} utilize only angular and energy information on a per-hit basis for cluster reconstruction. In contrast, \textit{Agglomerative Clustering}, \textit{Edge Clustering (with time)}, and \textit{Agglo + Edge (with time)} additionally incorporate time-of-hit information into the clustering process. Uncertainties reported for the four edge detection neural network variants correspond to the standard deviation of the results obtained from ten independent training runs.}
\label{tab:results}
\end{table}
The results of this study are summarized in Table~\ref{tab:results}, organized according to increasing levels of reconstruction complexity. For completeness, the previously obtained results from the comparison between the "baseline"geometrical R3B clustering algorithm and the agglomerative model are also included.\newline
The agglomerative model shows improved performance over the R3B baseline in terms of both event-level true positives (TP) and cluster-level (WR) values. However, it exhibits inferior performance with respect to the false negative (FN) rate, indicating a tendency to miss relevant hits during reconstruction. This limitation motivated the development of an Edge Detection Neural Network, initially evaluated as a standalone clustering algorithm and subsequently integrated into the agglomerative framework, yielding the combined model denoted as \textit{Agglo + Edge}.\newline
The \textit{Agglo + Edge} model demonstrates superior performance across all evaluated metrics, achieving an overall correct reconstruction rate of 81.3\%, significantly outperforming the \textit{Geometrical R3B Clustering} algorithm, which reaches 60.6\%.\newline
A visual representation of an example event, contrasting the incorrectly merged hits from the geometrical R3B clustering with the correctly reconstructed clustering using the Agglo+Edge model, is shown in Figure~\ref{fig:comparison_r3b_agglo_edge}.\newline

To further explore the capabilities of neural network-based clustering approaches, two additional models were evaluated: a standalone \textit{Edge Detection Neural Network} and a hybrid approach combining \textit{Geometrical R3B Clustering} with edge-based postprocessing (\textit{R3B + Edge}). Notably, both of these models operate without incorporating time-of-hit information, similar to the R3B baseline. Nonetheless, both outperform the \textit{Geometrical R3B Clustering}, underscoring the potential of edge-based neural network models for improving cluster reconstruction in high-granularity detector systems.\newline
The edge detection NNs presented here represent a special case of Graph Neural Networks (GNNs)~\cite{battaglia2018relational}, which, along with the more sophisticated transformer models~\cite{vaswani2017attention,amatriain2023transformer}, have seen widespread adoption in particle physics over the past five years~\cite{dezoort2021charged,ju2021performance,van2024transformers}. Interestingly, for this application, using an unsupervised learning algorithm (agglomerative clustering) to first define a graph structure presented a powerful inductive bias for our application which much improved our results over the standalone edge-NN.

\section{Outlook}\label{sec:disc_outlook}

The results presented in the previous section clearly demonstrate that high-level machine learning approaches, such as the Edge Detection NN, can significantly enhance the accuracy of cluster reconstruction. These models not only reduce distortions in the measurement process but also exhibit increased sensitivity to low-statistics reactions -- an important feature for experiments targeting rare processes.\newline
It is noteworthy that even the models, which do not utilize time-of-hit information (similarly to the \textit{Geometrical R3B Clustering}), outperform the baseline method. This underscores the general effectiveness of neural network-based methods in extracting structural features from detector data.\newline
The inclusion of time-of-hit information proves to be a critical factor for enhancing clustering performance. As this observable is typically available for CALIFA at R3B, the results of this study support the recommendation to incorporate it into the reconstruction pipeline wherever possible.\newline
Furthermore, these findings are intended to encourage broader adoption of advanced machine learning techniques by experimental groups, particularly in setups involving highly granular detectors. Such tools offer substantial performance benefits and can support more precise event reconstruction.\newline
One inherent limitation of the applied approach is its inability to correct for overly aggressive pre-clustering. In particular, false positive assignments introduced during the initial stage cannot be mitigated during the subsequent clean-up step by the edge-NN. This limitation is visible in Fig.~\ref{fig:2_1_mev_r3b_aggloNN}, where a slight excess of reconstructed counts at \(E_{reco} \approx 6.3\ \mathrm{MeV}\) is observed, likely indicating erroneous merging of unrelated hits due to excessive clustering. Despite this artifact, the high false negative rate -- exceeding the false positive rate by more than a factor of five in the baseline R3B clustering (see Table~\ref{tab:results}) -- motivated the development of a clustering strategy that prioritizes the recombination of hits to form complete clusters.\newline
Subsequent work could consider also adding a subsequent cluster splitting step in an end-to-end optimizable algorithm. Although, in principle, transformers could learn the graph structure directly from hit distributions, initial tests showed limited performance, highlighting an opportunity for the community to further develop combined machine learning-based reconstruction methods.\newline
From a computational standpoint, both the geometrical R3B clustering and the agglomerative clustering algorithms scale quadratically with the number of input hits, exhibiting a time complexity of \(\mathcal{O}(N^2)\), where \(N\) denotes the number of detector hits per event. The combined methods --R3B + Edge and Agglo + Edge -- induce additional computational overhead due to the Edge Detection Neural Network (NN) employed in the second stage. The current network architecture comprises three fully connected hidden layers with up to \(10^3\) neurons each, resulting in large matrix operations that dominate the runtime for typical events with \(N \sim \mathcal{O}(10^2)\). Consequently, future work will focus on optimizing the Edge Detection NN by significantly reducing the model size to enable faster execution while improving performance compared to the conventional geometrical R3B clustering.\newline
Additionally, transformer-based models \cite{vaswani2017attention} -- capable of analyzing full event topologies -- may offer further improvements in clustering accuracy by capturing complex, global features.

\section*{Acknowledgements}
The work was supported by BMBF 05P24WO2 and Excellence Cluster ORIGINS from the DFG (Excellence Strategy EXC-2094-390783311). It was made possible through the close collaboration of experts from different disciplines within the Cluster of Excellence ORIGINS~\cite{origins2025}.




\bibliographystyle{elsarticle-num} 
\bibliography{literature.bib}






\end{document}